\documentclass[aps,prb,preprint,superscriptaddress]{revtex4}

\usepackage{graphicx}% Include figure files
\usepackage{dcolumn}% Align table columns on decimal point
\usepackage{bm}% bold math

\begin{document}

\title{Unusual metamagnetism in CeIrIn$_5$}

\author{C. Capan}
\affiliation {Department of Physics and Astronomy, University of
California Irvine, Irvine, CA 92697-4575}\affiliation{Department of
Physics and Astronomy, Louisiana State University, Baton Rouge, LA
70803}
 \author{L. Balicas}
\affiliation{National High Magnetic Field Laboratory, Florida State
University, Tallahassee, Florida 32310}
\author{T. P. Murphy}
\affiliation{National High Magnetic Field Laboratory, Florida State
University, Tallahassee, Florida 32310}
\author{E. C. Palm}
\affiliation{National High Magnetic Field Laboratory, Florida State
University, Tallahassee, Florida 32310}
\author{R. Movshovich}
 \affiliation{Los Alamos National Laboratory,
MST-10, Los Alamos, New Mexico 87545}
\author{D. Hall}
\altaffiliation[Present affiliation: ]
                     {Physical Review Letters}
\affiliation{National High Magnetic Field Laboratory, Florida State
University, Tallahassee, Florida 32310}
\author{S. W. Tozer}
\affiliation{National High Magnetic Field Laboratory, Florida State
University, Tallahassee, Florida 32310}
\author{M. F. Hundley}
\affiliation{Los Alamos National Laboratory, MST-10, Los Alamos, New
Mexico 87545}
 \author{E. D. Bauer}
 \affiliation{Los Alamos National Laboratory,
MST-10, Los Alamos, New Mexico 87545}
 \author{J. D. Thompson}
\affiliation{Los Alamos National Laboratory, MST-10, Los Alamos, New
Mexico 87545}
 \author{J. L. Sarrao}
\affiliation{Los Alamos National Laboratory, MST-10, Los Alamos, New
Mexico 87545}
\author{J. F. DiTusa}
\affiliation {Department of Physics and Astronomy, Louisiana State
University, Baton Rouge, LA 70803}
\author{R. G. Goodrich}
\affiliation{Department of Physics, George Washington University,
Washington, DC 20052} \affiliation{Department of Physics and
Astronomy, Louisiana State University, Baton Rouge, LA 70803}
\author{Z.Fisk}
\affiliation {Department of Physics and Astronomy, University of
California Irvine, Irvine, CA 92697-4575}
\date{\today}

\begin{abstract}
We report a high field investigation (up to 45 T) of the
metamagnetic transition in CeIrIn$_5$ with resistivity and
de-Haas-van-Alphen (dHvA) effect measurements in the temperature
range 0.03-1 K. As the magnetic field is increased the resistivity
increases, reaches a maximum at the metamagnetic critical field, and
falls precipitously for fields just above the transition, while the
amplitude of all measurable dHvA frequencies are significantly
attenuated near the metamagnetic critical field. However, the dHvA
frequencies and cyclotron masses are not substantially altered by
the transition. In the low field state, the resistivity is observed
to increase toward low temperatures in a singular fashion, a
behavior that is rapidly suppressed above the transition. Instead,
in the high field state, the resistivity monotonically increases
with temperature with a dependence that is more singular than the
iconic Fermi-liquid, temperature-squared, behavior. Both the damping
of the dHvA amplitudes and the increased resistivity near the
metamagnetic critical field indicate an increased scattering rate
for charge carriers consistent with critical fluctuation scattering
in proximity to a phase transition. The dHvA amplitudes do not
uniformly recover above the critical field, with some hole-like
orbits being entirely suppressed at high fields. These changes,
taken as a whole, suggest that the metamagnetic transition in
CeIrIn$_5$ is associated with the polarization and localization of
the heaviest of quasiparticles on the hole-like Fermi surface.
\end{abstract}


\maketitle

%introduction
\indent Itinerant Electron Metamagnetism (IEM) refers to a field
induced transition from a paramagnetic (PM) to a field-polarized
paramagnetic state, first predicted for exchange-enhanced
paramagnets such as Pd \cite{WR1962}. The effect is intriguing since
large, almost discontinuous changes in the magnetization can be
observed through such a transition. In fact such field-induced
magnetization jumps have usually been interpreted as a dramatic
change in the density of states (DOS) at the Fermi energy as a
Zeeman split DOS peak crosses the Fermi level. The most spectacular
examples of IEM have been observed in the MgCu$_2$ type cubic Laves
phases such as \emph{A}Co$_2$ (\emph{A}$=$Sc, Y, Lu) \cite{Laves}.
In heavy fermion systems, the screening of local moments by
conduction electrons leads to a Kondo resonance peak in the DOS
close to the Fermi level, so that large magnetic fields can lead to
IEM behavior. Among the first heavy fermion compounds reported to
have a metamagnetic transition are UCoAl \cite{UCoAl}, and UPt$_3$
\cite{UPt3}, and, by far the most thoroughly investigated,
CeRu$_2$Si$_2$ \cite{Flouquet}. Theoretical approaches that take
into account the spin fluctuations can successfully capture the
basic features of IEM in Laves phase compounds\cite{Yamada}. In
order to capture some of the more subtle aspects of IEM in heavy
fermion systems, Hubbard or Anderson model physics must be included.
These models can reproduce features such as the Fermi surface volume
change inferred from the dHvA \cite{Aoki} and due to $f$-electron
localization\cite{Meyer}, as well as the sign change of the exchange
coupling \cite{Satoh} (from antiferro- to ferromagnetic) observed in
neutron scattering \cite{Rossat-Mignod}.

\indent Despite a great amount of both theoretical and experimental
work, some of the essential questions about IEM remain unsolved. For
instance, is there a common picture emerging between the \emph{d}-
and \emph{f}-electron systems that exhibit IEM? Would it involve
solely a change in Fermi surface geometry (along with a
magnetoelastic instability) or include a change in the volume
enclosed by the Fermi surface? This is a particularly relevant issue
for the heavy fermion systems, since Fermi surface changes have been
observed under applied field\cite{CeIn3-Harrison,YRS-2008} and
pressure\cite{shishido-cerhin5} and it is not clear if this
indicates $f$-electron localization due to field polarization of the
conducting electrons. Finally the discovery of IEM in the
Ruddlesden-Popper compounds Sr$_3$Ru$_2$O$_7$ and
Sr$_4$Ru$_3$O$_{10}$ has opened new horizons \cite{Sr327-Chiao,
Sr327-Science, Mao2006}. Systematic deviations from the Fermi Liquid
(FL) theory are reported in the resistivity ($\rho$)
\cite{Sr327-Science}, thermal expansion \cite{Sr327-thermal}, and
NMR \cite{Sr327-NMR} near the metamagnetic transition in
Sr$_3$Ru$_2$O$_7$, raising the possibility of a field-induced
quantum critical point (QCP) in IEM systems. The $T=0$ suppression
of a first order metamagnetic discontinuity terminating at a
critical end-point has been shown theoretically to result in a QCP
\cite{Andy}, a situation potentially realized in Sr$_3$Ru$_2$O$_7$.
The role of disorder \cite{Sr327-upturn, Sr327-science07} as well as
the formation of magnetic domains \cite{Binz-Condon} in the
proximity of a metamagnetic transition are still under
investigation.

\indent Here we report a high field investigation of the recently
discovered IEM transition in CeIrIn$_5$ \cite{Takeuchi,Palm,
Stewart}, to search for evidence of a metamagnetic QCP. CeIrIn$_5$
is a heavy fermion superconducting member of the layered tetragonal
CeMIn$_5$ (M=Rh,Ir,Co) family \cite{Petrovic}. Since their
discovery, the 1-1-5 compounds have been the focus of much attention
because they exhibit striking deviations from FL theory. There has
been a great deal of work trying to understand the origin of such
anomalous behavior in strongly correlated systems \cite{Stewart-RMP}
and descriptions are often based on the proximity to a QCP in their
phase diagrams.  Near such a singular point where a magnetic phase
transition is driven to $T=0$ by an external tuning parameter
(pressure, magnetic field or doping) the presence of quantum
fluctuations is expected to have a strong effect on all physical
properties, leading to the breakdown of the FL theory. In
particular, the 1-1-5 compounds have rich and complex phase
diagrams, with notably a field tuned QCP near the superconducting
upper critical field in CeCoIn$_5$ \cite{CeCoIn5JP, CeCoIn5Andrea,
CeCoIn5Filip} and a pressure tuned QCP in CeRhIn$_5$
\cite{shishido-cerhin5, CeRhIn5Tuson}. Besides their phase diagram,
their heavy fermion state has also attracted attention in its own
right. A careful analysis of the transport and thermodynamic
properties of La-diluted CeCoIn$_5$ and CeIrIn$_5$ suggests that the
coherent heavy fermion ground state coexists with a finite fraction
of single-ion Kondo centers down to the lowest $T$ \cite{twofluid}.
This conclusion was further supported by optical conductivity
measurements on CeMIn$_5$ samples that revealed a broad
hybridization between the conduction electrons and the more
localized $f$-electrons which included contributions from 4
electronic bands\cite{burch}. In addition, the hybridization gaps
that result appear to have extensive momentum dependence with
regions where the smallest gap, associated with a particular hole
band ($\beta $ orbits in de Haas-van Alphen experiments), is seen to
go to zero. The conclusion was that Kondo screening could be
incomplete, particularly in regions where the hybridization gap goes
to zero. Band structure calculations using DMFT, motivated by the
optical conductivity measurements, have found two distinct
hybridization gaps in CeIrIn$_5$, attributed to two distinct In
sub-bands hybridizing with the Ce lattice\cite{dmft}.

\indent In addition to the many common trends with its Co and Rh
counterparts, a distinguishing feature of CeIrIn$_5$ is the presence
of a metamagnetic-like transition near 30 T for the [001] field
orientation (perpendicular to the CeIn$_3$ planes). This transition
is characterized by a non-linear increase in magnetization
\cite{Takeuchi,Palm} and a $\lambda$-like anomaly in the specific
heat \cite{Stewart}. Because the family of 1-1-5 compounds has many
unusual features, including what appear to be unusual transitions
perhaps associated with interesting QCPs, we are compelled to
explore the possible existence of a metamagnetic QCP in CeIrIn$_5$.
Our earlier specific heat and $\rho$ measurements up to 17 T have
shown a field-induced non-Fermi Liquid behavior (NFL). However the
limited field range of these experiments did not allow us to explore
the immediate vicinity of the metamagnetic transition
\cite{capan:180502}. We have subsequently extended our
investigations to higher fields (up to 45 T) and explored the angle
dependence of the high field transition in CeIrIn$_5$ combining
resistivity and torque magnetometry measurements\cite{SCES07}. Here,
we give a complete account of our high field investigations with an
in-depth analysis in search for Fermi surface changes at the
transition and for clues about the cause of the NFL behavior.

\indent Our measurements reveal several changes that occur at or near
the metamagnetic critical field which have important consequences for
interpreting the mechanism for IEM in CeIrIn$_5$.  The most obvious is
that an increased scattering rate for charge carriers results in an
increased resistivity and a decreased dHvA amplitude in the region of
the transition. This is accompanied by a resistivity anomaly at very
low temperature that switches from increasing resistivity with
decreasing temperature below the transition, to decreasing resistivity
with decreasing temperature above the transition, none of which can be
interpreted as a Fermi liquid-like behavior. Finally, we observe the
complete suppression of two of the heaviest dHvA orbits in the high
field state, while other orbits recover and eventually exceed their
lower field amplitude. These changes are \emph{not} accompanied by
significant variations in either the dHvA frequency, indicating no
measurable change to the extremal areas of the Fermi surface, or the
cyclotron masses associated with orbits that we observe. Our data
indicate that light parts of the Fermi surface are left unchanged by
the IEM while the heavier parts are suppressed beyond the metamagnetic
critical field. Although our data do not provide direct evidence for a
metamagnetic QCP, we observe NFL transport properties in proximity to
the critical field as well as an instability of the heaviest Fermi
surface sheet, both of which are consistent with quantum critical
behavior.

\indent The paper is organized as follows: we first present the high
field phase diagram of CeIrIn$_5$, which has been previously
established via magnetization \cite{Takeuchi} and specific heat
measurements\cite{Stewart}. This is followed in section III by a
description of the resistivity anomalies associated with the
metamagnetic transition.  Section IV is devoted to the evolution of
the Fermi surface across the metamagnetic transition, with an
analysis of the dHvA oscillations observed in the torque signal.
Finally, in section V we summarize the conclusions that our data
make necessary.

\section{Experimental Details}

\indent Isothermal magnetoresistance sweeps have been performed
using standard four probe techniques and a low frequency resistance
bridge, on two single crystals at the National High Magnetic Field
Laboratory, using the 33 T resistive magnet for the first and the 45
T hybrid magnet for the second sample. The samples were pre-screened
for In inclusions (by checking for superconducting signal in the
resistivity at the $T_c$ of In) and mounted on a rotating sample
holder inside the mixing chamber of a top loading dilution
refrigerator. The sample $S1$ was measured with a current of 300
$\mu$A up to 33 T at a fixed angle of $14^o$ with respect to [001].
These data were later complemented by measurements up to 45 T on
sample $S2$ with varying angle and a current of 1 mA. For both
samples, the current is applied perpendicular to [001]. The field
was rotated from [001] to a direction parallel to the ab-plane, but
perpendicular to the applied current. The samples were allowed to
thermalize for about 1 hr and the temperature of the mixing chamber
was recorded at 0 T ($S1$) and 12 T ($S2$) before each field sweep.
Simultaneous torque magnetization measurements have been performed
using an $AC$ capacitance bridge on a sample cut from the first
crystal ($S1$) and mounted with G.E. varnish on a Cu-Be cantilever.

\section{Phase Diagram}

\begin{figure}
\resizebox{!}{0.8\textwidth}{\includegraphics{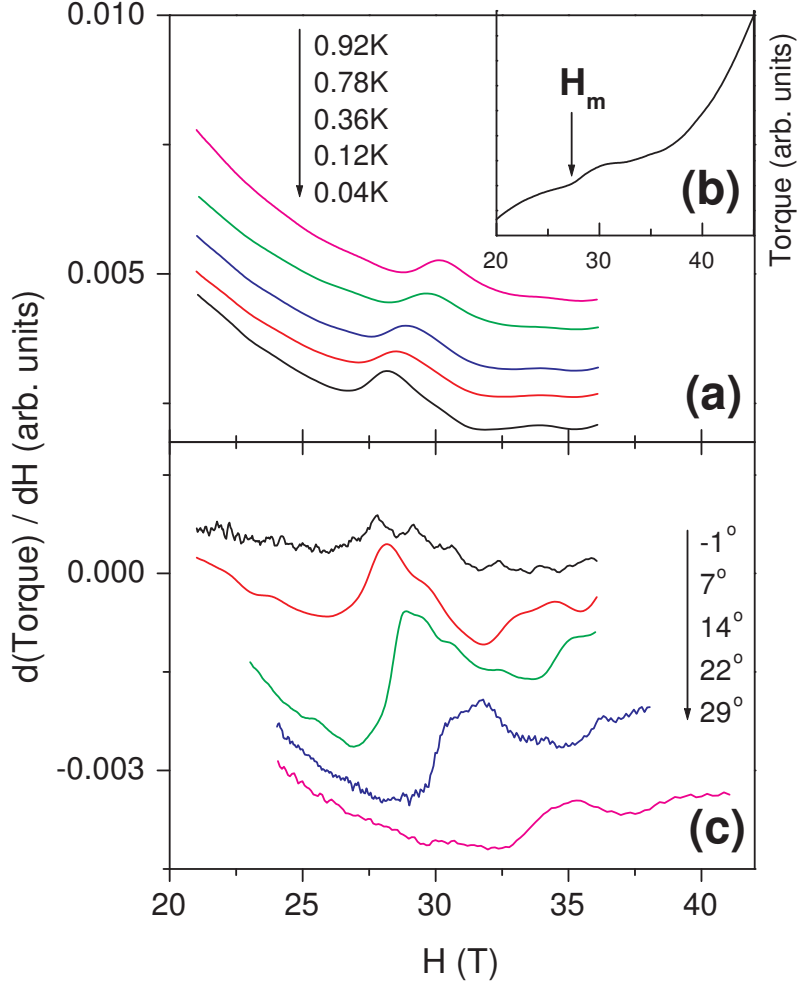}}
\caption{\label{Fig1}  Torque signal near the metamagnetic critical
field. (a) Derivative of Torque signal vs. applied magnetic field,
$H$, in CeIrIn$_{5}$ in the field range 20-35 T at an angle of
$-8^{\circ}$ for temperatures ranging between 0.04 and 0.92 K.  The
peak corresponds to the metamagnetic transition. (b) Torque signal
at 0.04 K (in arbitrary units) vs.~$H$ between 20 and 45 T for
$\Theta=-8^{\circ}$. The metamagnetic field $H_m$ is indicated by an
arrow. (c) Derivative of the Torque signal vs.~$H$ for field
orientations ranging between $-1^{\circ}$ and $29^{\circ}$ at the
base temperature of 0.04 K. The data are shifted vertically for
clarity.}
\end{figure}

\begin{figure}
\resizebox{!}{0.8\textwidth}{\includegraphics{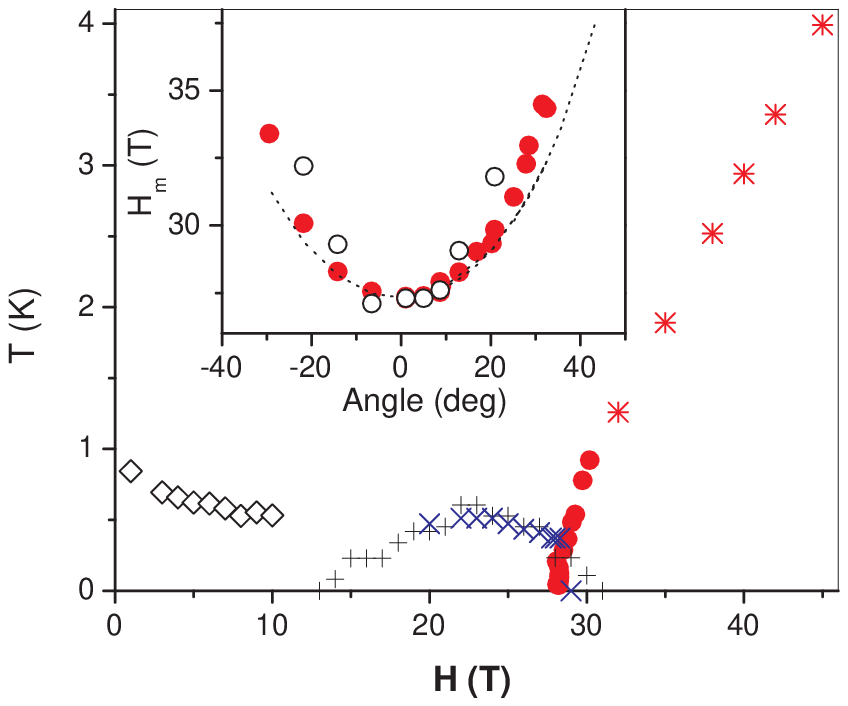}}
\caption{\label{Fig2} Magnetic field phase diagram in CeIrIn$_5$ for
$H\| [001]$. The metamagnetic transition $H_{m}$ is determined from
torque magnetometry ($\bullet$) and from specific heat data of
Ref.~\cite{Stewart} ($\ast$). The Fermi Liquid crossover temperature
is determined from our previous specific heat
data\cite{capan:180502} ($\diamond$). The resistivity minimum for
the two samples $S1$ ($\times$) and $S2$ ($+$) are also shown.
Inset: Angular dependence of the metamagnetic field $H_{m}$
determined from magnetization ($\bullet$) and from resistivity
($\circ$) at 0.04 K. The angle ($\Theta$) is defined between the
magnetic field and [001]. The dashed line correspond to a fit to
$H_m = H_m(0) \frac{1}{cos(\Theta)}$.}
\end{figure}

\indent The high field phase diagram has been previously established
by magnetization \cite{Takeuchi} and specific heat \cite{Stewart}
measurements. Figures \ref{Fig1}a and \ref{Fig1}c show the
derivative of the torque signal as a function of magnetic field,
$H$, for temperatures ranging between 0.04 and 0.92 K at fixed angle
($-8^{\circ}$) and, in Fig.~\ref{Fig1}c, for angles ($\Theta$)
between $H$ and [001] ranging from $-1^{\circ}$ to $29^{\circ}$ at
our base temperature (0.04 K).  The torque signal is proportional to
the vector product of magnetization with magnetic field and is shown
for $\Theta=-8^{o}$ and $T=0.04$ K as an inset (see
Fig.~\ref{Fig1}b). The torque signal is reported in Fig.~\ref{Fig1}
in arbitrary units since the cantilever used was uncalibrated. The
feature in the torque signal, indicated by an arrow in figure
\ref{Fig1}b , or equivalently the peak in the derivative,
Fig.~\ref{Fig1}a and \ref{Fig1}c, marks the metamagnetic transition
field $H_m$. $H_m$ is seen to shift to higher fields with increasing
temperatures and angles. On the $H-T$ phase diagram, shown in
Fig.~\ref{Fig2}, the transition determined from torque data (at
$\Theta = -8^{\circ}$ from [001]) extends to lower temperatures the
previously reported transition line determined from the specific
heat\cite{Stewart}. The transition field $H_m$ increases nearly
quadratically in temperature. This dependence of $H_m(T)$ is also
observed in the Laves compounds and is consistent with
Ginzburg-Landau theory\cite{Yamada}. The angular dependence of
$H_m$, shown in the inset of Fig.~\ref{Fig2}, can be fit to a
$\frac{1}{cos(\Theta)}$ dependence, suggesting that only the
component of the field parallel to [001] induces the transition. In
fact no transition is observed up to 60 T for fields along the
ab-plane \cite{Takeuchi}. The transition is most likely second order
in the phase space of temperature, field and angle that we explored,
with no sharpening of magnetization anomaly into a first-order like
discontinuity upon cooling, in contrast to the previous claim of
Ref.\cite{Stewart}. Although we cannot exclude a weakly first order
transition, the torque data do not show any evidence for a critical
end-point in CeIrIn$_5$, as is the case for Sr$_3$Ru$_2$O$_7$
\cite{Sr327-Science}. The transition can be extrapolated to $T=0$
around $H_m(\emph{T}=0)=28$ T. Thus the phase diagram suggests the
possibility of a metamagnetic QCP for $ H \parallel [001]$. Also
shown in the phase diagram is the crossover line from Fermi Liquid
to non-Fermi Liquid regime obtained in our earlier specific heat
$(C)$ investigation (up to 17 T) as the temperature below which
$\frac{C}{T}$ becomes constant.

\section{Magnetoresistance}

\begin{figure}
\resizebox{!}{0.8\textwidth}{\includegraphics{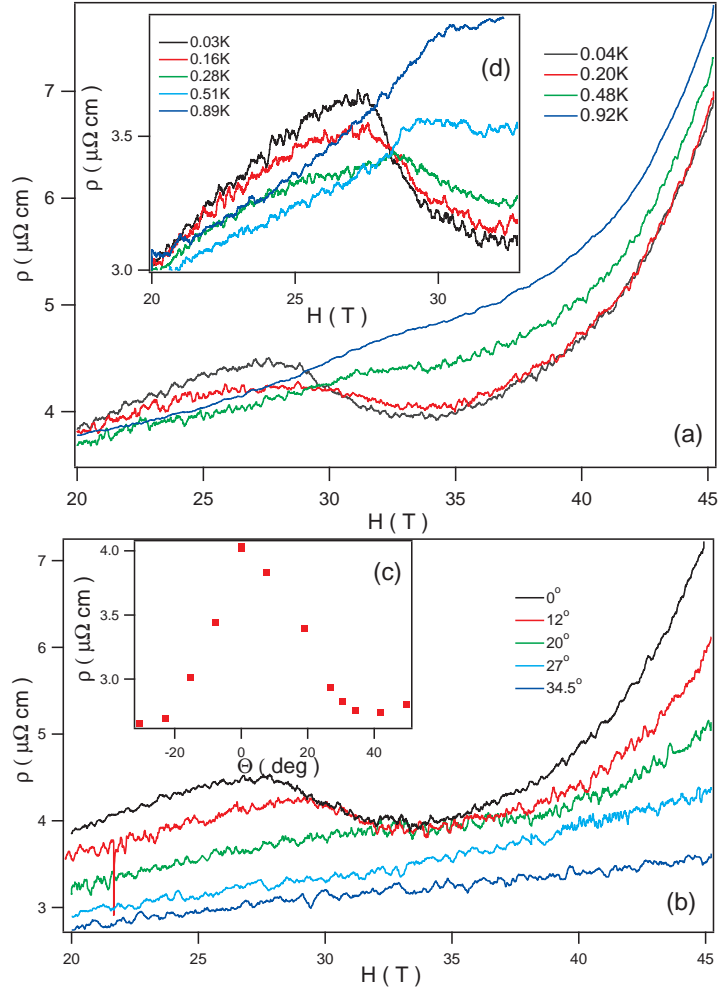}}
\caption{\label{Fig3} Magnetoresistance in CeIrIn$_5$. Resistivity
$\rho$ vs applied magnetic field $H$ in CeIrIn$_5$ for the sample
$S2$ measured up to 45 T (a) for indicated temperatures at
$\Theta=-8^{o}$ (b) for various angles $\Theta$ between the field
and the [001] orientation at the base temperature of $40$ mK.
Insets: (c) Angular dependence of $\rho$ at $T=40$ mK and $H=20$ T
in sample $S2$. (d) Resistivity vs applied magnetic field for the
sample $S1$, measured up to 33 T, for indicated temperatures, at
fixed angle $\Theta=14^{o}$. }
\end{figure}

The transition observed in $M(H)$ is also apparent in the
resistivity data shown in Fig.~\ref{Fig3}. The magnetoresistance
(MR) in CeIrIn$_5$ is positive at low $T$, except around $H_m$.
Indeed, $\rho$ goes through a broad maximum and decreases above
$H_m$ for fields oriented close to [001] ($\Theta \leq 12^o$)
(Fig.~\ref{Fig3}b). With a further increase of $H$, $\rho$ increases
steeply in the high field state. The field at which $\rho$ is
maximum closely follows the same temperature and angular dependence
as $H_m$ determined from magnetization, and is also shown in the
inset of Fig.~\ref{Fig2}.  Note that a crossover from positive to
negative MR has been reported in the related compound CeCoIn$_5$
\cite{CeCoIn5JP}, however, its relation to a possible metamagnetic
transition is not well established. With increasing $\Theta$ the
peak shifts to higher fields and becomes less pronounced: at
$\Theta=20^{\circ}$ a plateau is observed. A small change in the
slope can still be resolved at $\Theta=27^{\circ}$ but no hint of a
transition is observed at $\Theta=34.5^{\circ}$.

\begin{figure}
\resizebox{!}{0.8\textwidth}{\includegraphics{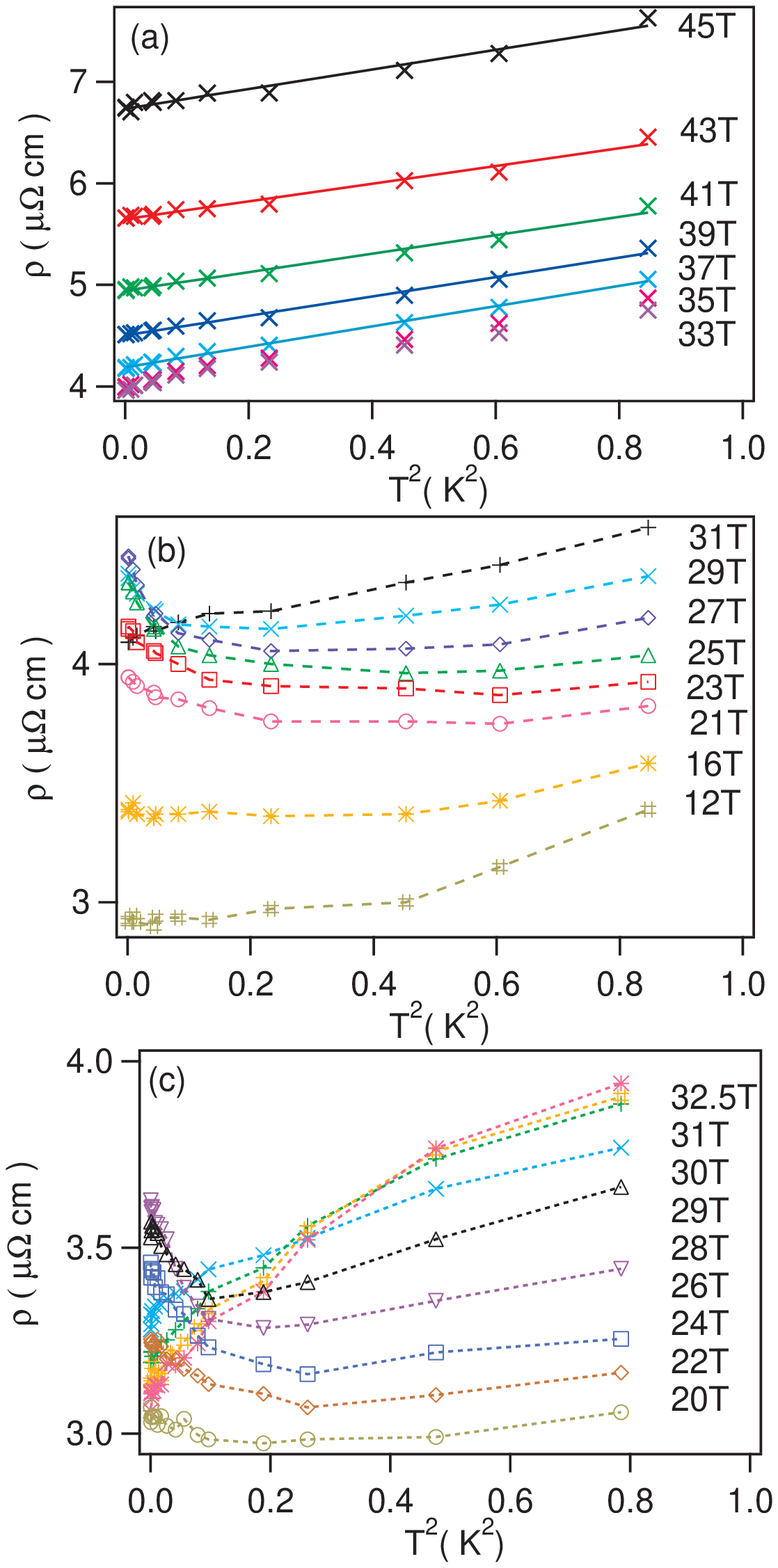}}
\caption{\label{Fig4} Temperature dependence of resistivity in
CeIrIn$_5$ for a field applied at $\Theta=-8^{o}$ from [001].
Resistivity, $\rho$, vs temperature, $T$, squared (a),(b) for the
sample $S2$ between $12$ and $45$ T in the $T$-range $0.04-0.92$ K
and (c) for the sample $S1$ between $20$ and $32.5$ T in the
$T$-range $0.03-0.87$ K. Solid lines in (a) are linear fits to
$\rho$ vs. $T^2$.}
\end{figure}

\indent The temperature dependence of the resistivity is constructed
from the isothermal sweeps, Fig.~\ref{Fig3}, and is shown for a
range of fields in Fig.~\ref{Fig4} where $\rho$ is plotted as a
function of $T^2$. The resistivity is consistent with a $T^2$
dependence between 0.04 K and 0.92 K at fields larger than 36 T
where the sample has entered the field-polarized paramagnetic state.
This is as expected for the case of a Fermi liquid ground state.
Fits to a $T^2$ dependence to our data are shown as solid lines in
Fig.~\ref{Fig4}a for fields between 37 and 45 T. The slope in
$\rho=\rho_{0} + A_{FL} T^{2}$ defines the Fermi Liquid coefficient
$A_{FL}$, with $A_{FL} \sim 0.972$ $\mu \Omega$ cm/K$^2$ (at $H=45$
T) corresponding to a specific heat coefficient of $\gamma \sim
0.54$ J/K$^{2}$ mol (using the Kadowaki-Woods ratio obtained for
CeCoIn$_5$ in Ref.~\cite{CeCoIn5Andrea}), which is close to the
experimental value found in CeIrIn$_5$ at $H=45$ T ($\gamma \sim
0.4$ J/K$^{2}$ mol in Ref.~\cite{Stewart}), and slightly suppressed
compared to $\gamma \sim 0.75$ J/K$^{2}$ mol at $H=1$
T\cite{capan:180502}, as expected. This estimate suggests that the
field-polarized state is still a heavy fermion metal, a conclusion
also supported by the analysis of the cyclotron effective masses
(see below).

\indent In contrast, for fields just above $H_m$ there is a strong
deviation from a $T^2$ dependence of $\rho$ at the lowest
temperatures. Here the resistivity has a temperature dependence
weaker than quadratic, as evidenced by the downward curvature in the
figure for sample $S2$ (Fig.~\ref{Fig4}b) between 33 and 35 T, and
for sample $S1$ between 30 and 32.5 T (Fig.~\ref{Fig4}c). This non
quadratic T-dependence of $\rho$ occurs very close to the critical
field, $H_m$, for both $S1$ and $S2$. In a previous publication,
where we measured a different CeIrIn$_5$ sample, we found that
$\rho$ was not well described by a Fermi liquid-like $T^2$
dependence for temperatures down to 50 mK and magnetic fields of 12,
15, and 17 T oriented parallel to the [001] direction whereas a
$T^2$ dependence was found with the field oriented perpendicular to
[001]\cite{capan:180502}. This stresses the importance of
magnetoresistance effects in the high field regime for the
$T-$dependence of $\rho$, which can make the interpretation of
non-Fermi Liquid behavior problematic. However, the anomalous
temperature dependence of $\rho$ in CeIrIn$_5$ is clearly associated
with the metamagnetic transition since the quadratic-in-$T$ behavior
is recovered for fields well above $H_m$. Note that the occurrence
of non Fermi-liquid like behavior in the resistivity is somewhat at
odds with the simultaneous observation of quantum oscillations. This
suggests that the inelastic (electron-electron) scattering
responsible for the former becomes negligible compared to the
elastic (disorder) scattering at very low T. This situation is
\emph{not} unique to CeIrIn$_5$: CeCoIn$_5$ also exhibits non-Fermi
Liquid behavior concomitant to quantum oscillations close to its
superconducting upper critical field\cite{CeCoIn5JP,CeCoIn5Andrea,
CeCoIn5Filip,CeCoIn5-dHvA}. Our observation of a non-quadratic
temperature dependence of $\rho$ near $H_m$ is in direct conflict
with the observation of a resistivity well described by a $T^2$
dependence across the metamagnetic transition in CeRu$_2$Si$_2$
\cite{CeRu2Si2-SJulian}, suggesting a separate mechanism for IEM in
these two heavy fermion systems.

\indent For fields below $H_m$, a new contribution to $\rho$ appears
with a negative $d\rho/dT$ below $\sim$ 0.5 K, as seen in
Fig.~\ref{Fig4}. This low-$T$ upturn in $\rho$ becomes more pronounced
in the vicinity of the metamagnetic field, and is rapidly suppressed
above it. The temperature $T_{\text{min}}$ where $\rho(T)$ has a
minimum is field dependent and defines the dome-shaped region in the
$H-T$ phase diagram, as shown in Fig.~\ref{Fig2}. We note that there
is no anomaly associated with $T_{\text{min}}$ in the torque signal.
This behavior has been observed in two different crystals at two
different field orientations close to [001], and so it is a robust and
reproducible feature of the metamagnetic transition in CeIrIn$_5$. \\

\indent An upturn in $\rho$ in the presence of high magnetic field
is also observed at low$-T$ in CeCoIn$_5$
\cite{CeCoIn5JP,CeCoIn5Filip} and UPt$_3$ \cite{UPt3upturn}. This
contribution becomes more pronounced as the field is increased. It
has been so far associated with the cyclotron magnetoresistance in
the high field limit ($\omega_c \tau \sim $1), since it is only
observed in the transverse geometry. An upturn is indeed possible in
the high field limit of a compensated metal with $\rho(H,T) \sim
\frac{H^2}{(\rho_0 + A T^2)^2}$. The upturn in CeIrIn$_5$ is
saturating at low $T$ and can be reasonably well described by this
form. However, its suppression above the metamagnetic field is at
odds with this scenario. This cannot be accounted for by the
negative contribution to MR compensating the positive one, since MR
is only negative close to $H_m$ and the upturn in $\rho$ is
\emph{not} recovered when the MR becomes positive above 36~T.
Moreover, the dome-shaped onset temperature $T_{\text{min}}$, as
shown in Fig.~\ref{Fig2}, clearly does not track the crossover to
the high field regime ($\omega_c \tau \sim $1), in striking contrast
to CeCoIn$_5$\cite{CeCoIn5JP,CeCoIn5Filip}. To summarize, although
MR effects in the high field regime may account for the upturn in
$\rho(T)$, the suppression of this feature in the field polarized
state, in the absence of Fermi Surface change(see below), suggests
that a different mechanism is involved in CeIrIn$_5$.\\

\indent In Sr$_3$Ru$_2$O$_7$, a similar, but more pronounced,
resistivity anomaly has been observed in the cleanest samples
\cite{Sr327-upturn} and has been tentatively assigned to the
condensation of a nematic phase surrounding the QCP
\cite{Sr327-science07}. This is considered as a strong possibility
in Sr$_3$Ru$_2$O$_7$ since there are first order-like boundaries
within which the resistivity anomaly occurs, making the idea of a
phase transition appealing.  The most likely candidate for such a
domain-wall scattering scenario is Condon-like domains
\cite{Binz-Condon}, and the first order nature of the transitions
makes the inhomogeneous state hypothesis a sensible one. Despite
similar residual resistivities, the transition at $H_m$ in
CeIrIn$_5$ is not as sharp as the one observed in Sr$_3$Ru$_2$O$_7$.
Although we cannot rule out the possibility of a common mechanism,
the absence of a sharp first order boundary make the domain wall
scenario unlikely in CeIrIn$_5$.

\indent An alternative explanation for the low-$T$ resistivity
anomaly near $H_m$ in CeIrIn$_5$ relies on the scattering of
quasiparticles from quantum magnetic fluctuations leading to a
crossover into a diffusive transport regime. This idea is consistent
with the prediction that any deviation from Fermi Liquid behavior in
presence of disorder will lead to an infinite resistivity at $T=0$
\cite{Chandra}. Moreover, logarithmic corrections to the
conductivity, due to enhanced impurity scattering in the presence of
strong quantum fluctuations, have been predicted for a ferromagnetic
\cite{Belitz, Paul}, and more recently for a metamagnetic
\cite{YBK}, quantum critical point. The idea is that near the QCP,
the diverging magnetic coherence length (with decreasing $T$)
eventually becomes larger than the mean free path ($\ell$) (or
equivalently, the correlation time becomes larger than the time
between successive collisions) and the system effectively behaves as
if it were impurity scattering limited. This is similar to what was
previously reported in the itinerant ferromagnet
Fe$_{1-x}$Co$_{x}$Si\cite{Manyala}. The crossover from ballistic to
disordered regime is expected to occur near $T_0=\frac{\hbar
\tau^{-1}}{k_B} \sim 0.62$ K in CeIrIn$_5$ which is of the order of
$T_{\text{min}}$. Here we have used the Drude model along with the
measured linear temperature coefficient of the specific heat
($\gamma_0 = 0.75$ J mol$^{-1}$ K$^{-2}$ defined at $H=1$ T), along
with estimate of the electron density, 2.9 electron per Ce ion based
on the Hall coefficient at 300~K\cite{Hundley}, $n=17.98\times
10^{27}$ m$^{-3}$, in order to estimate the scattering time :
\begin{equation}\tau=\frac{m^{*}}{n e^{2} \rho} =
\frac{\gamma_{0}}{(\frac{\pi}{3})^{2/3}
(\frac{k_{B}}{\hbar})^{2}n^{4/3}e^{2}\rho},\label{eq:drude}\end{equation}
with the result that $\tau \sim 1.2~10^{-11}$s. However, as we lack
convincing evidence of a QCP at $H_m$ in CeIrIn$_5$, this hypothesis
must be viewed as only a possibility.

\indent In summary, we believe several of the suggested mechanisms
for the origin of the MR can be effectively ruled out, and one of
them, quantum corrections in a diffusive transport regime, depends
heavily on the existence of a QCP at $H_m$, which we do not have
conclusive evidence for at this time. In addition, the presence of
non-quadractic $T-$dependence in $\rho$, which is clearly related
with the metamagnetic transition in CeIrIn$_5$, distinguishes it
from the prototypical heavy fermion metamagnet CeRu$_2$Si$_2$. We
note that in the latter, a metamagnetic QCP has been recently ruled
out based on lower temperature, higher precision, resistivity
measurements \cite{CeRu2Si2-SJulian}. The low$-T$ upturn as well as
the non-quadratic behavior in $\rho$ raise important questions and
present a challenge
for understanding the unusual behavior that we observe in CeIrIn$_5$.\\

%dHvA results
\section{de-Haas-van-Alphen Effect}

The study of the Fermi surface in heavy fermion metals via quantum
oscillations of magnetization (dHvA effect) was pioneered by
Taillefer et.\ al \cite{UPt3dHvA}. These oscillations result from
the discontinuity in the Fermi-Dirac distribution at $T=0$ and are
well understood in simple metals in the framework of
Lifshitz-Kosevich (LK) theory \cite{Shoenberg}. The observation of
dHvA oscillations in heavy fermion metals is an important
experimental milestone showing that these strongly correlated
compounds can simply be described in terms of heavy quasiparticles,
Landau quasiparticles with cyclotron masses much larger than band
masses. The Onsager relation dictates that the dHvA frequencies are
proportional to the cross-sectional area(s) of the Fermi surface
perpendicular to the applied field. Therefore, by mapping the
angular dependence of the frequency one can determine the Fermi
Surface topography. This has been done in the 1-1-5 compounds, with
the observed quasi-2D $\alpha$ (electron) and $\beta$ (hole) FS
sheets in close agreement with band structure calculations
\cite{CeCoIn5-dHvA, Shishido, Ir115dHvA}. These results indicate
that both CeCoIn$_5$ and CeIrIn$_5$ have itinerant $f$-electrons at
low temperatures, whereas CeRhIn$_5$ has localized $f$-electrons.
Moreover, a band-by-band investigation of the mass enhancement is
possible with the dHvA technique, complementing the Fermi surface
averaged effective mass given by the specific heat coefficient. By
determining the angular dependence of the cyclotron mass it has been
possible to identify ``hot spots" in the parent cubic compound
CeIn$_3$\cite{ebihara}.\\

\begin{figure}
\resizebox{!}{0.8\textwidth}{\includegraphics{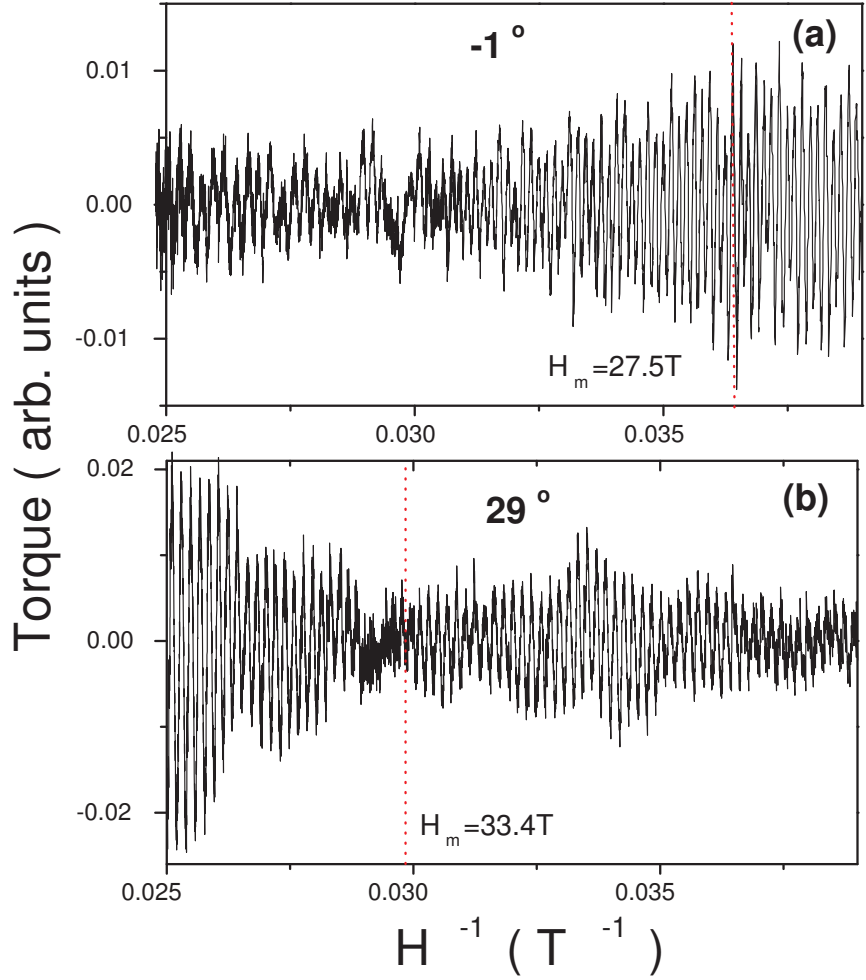}}
\caption{\label{Mosc} De Haas-van Alphen oscillations in CeIrIn$_5$.
Torque signal (after subtraction of a smooth background) vs inverse
magnetic field $H^{-1}$ at $\Theta=-1^{\circ}$ (a) and $29^{\circ}$
(b).  The corresponding field range is 25.6 to 40 T. The dashed line
indicates the metamagnetic transition field $H_m=$27.5 T and 33.4 T
for the two angles.}
\end{figure}

\begin{figure}
\resizebox{!}{0.8\textwidth}{\includegraphics{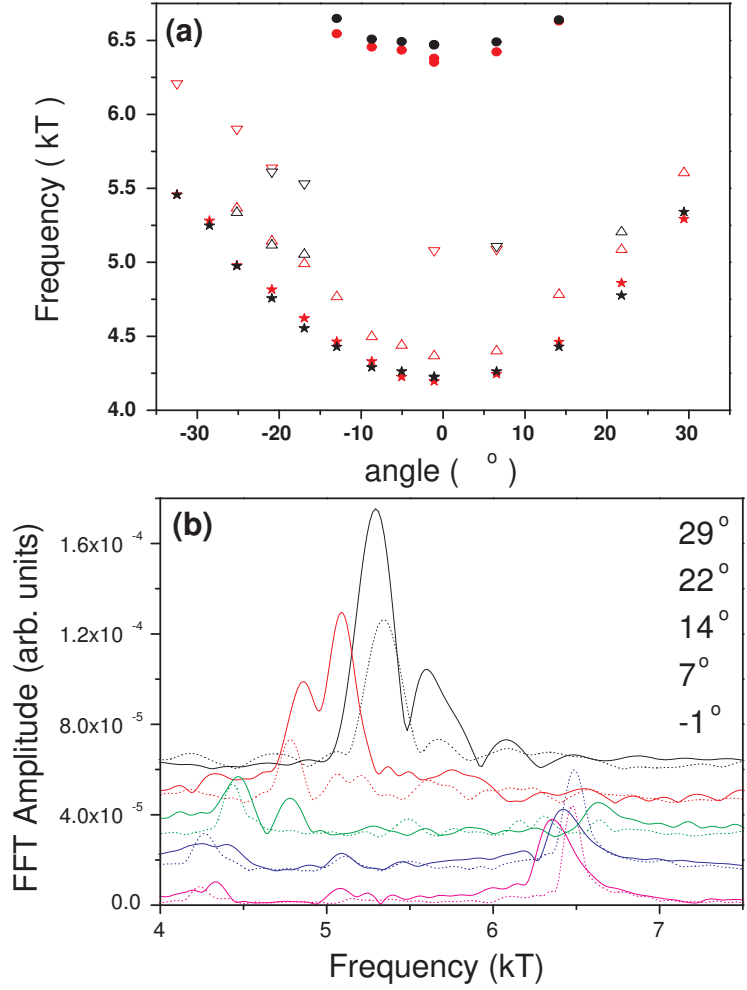}}
\caption{\label{FFTvsAngle} Angle dependence of the de Haas-van
Alphen frequencies in CeIrIn$_5$. (a) Frequency vs Angle in the low
field (black symbols, $H<H_{m}$) and high field states (red symbols,
$H>H_{m}$) corresponding to dHvA branches $\alpha_1$
($\bigtriangleup$), $\alpha_2$ ($\bigtriangledown$), $\alpha_3$
($\star$) and $\beta_2$ ($\bullet$). Zero angle corresponds to $H
\parallel $[001]. The $\beta_1$ orbit (at $10.4$ kT) is only observed
for $\Theta=14^{\circ}$ and is not shown in this figure. (b) Fast
Fourier Transform (FFT) spectra of the torque signal vs frequency in
the low field (dashed lines) and high field (solid lines) states for
field oriented at 29$^{\circ}$, 22$^{\circ}$, 14$^{\circ}$,
7$^{\circ}$ and -1$^{\circ}$ from [001] (top to bottom).}
\end{figure}

\indent Figure \ref{Mosc} shows the torque signal in CeIrIn$_5$ (in
arbitrary units) as a function of inverse magnetic field $H^{-1}$,
in the field range 25.6-40 T, at the base temperature 0.04 K, for
two extreme angles $-1^{\circ}$ and $29^{\circ}$. The background
field dependence (shown in Fig.~\ref{Fig1}b) was determined from a
$5^{th}$ order polynomial fit to the torque signal and then
subtracted from the data to better reveal the oscillations in the
torque signal which are periodic in $H^{-1}$ corresponding to the
dHvA effect. The dHvA frequencies are determined from the Fast
Fourier Transform (FFT) spectra of the torque signal at each angle,
shown in Fig.~\ref{FFTvsAngle}b. The FFT is performed on two field
intervals of equal length in $H^{-1}$ for $H<H_{m}$ and $H>H_{m}$,
thus keeping the same frequency resolution to allow comparison.
Fig.~\ref{FFTvsAngle}a compares these dHvA frequencies below and
above the metamagnetic transition on an angular interval of $\pm
30^{\circ}$ around [001]. The observed frequencies correspond to
extreme orbits on the $\alpha$ and $\beta$ sheets of the Fermi
surface, in agreement with previous reports (up to 16 T) in both
their value and their angular dependence. The lower
frequencies\cite{Shishido,Roy2009}, associated with the 3D electron
pockets, are not easily detected in torque magnetometry since the
torque signal is proportional to the angular derivative of the
frequency (see Ref.~\cite{Shoenberg}, p.87). Overall, the same
branches are observed both below and above H$_{m}$, with only a
slight shift in frequency across the metamagnetic transition.
Moreover, within our frequency resolution (set by the field range of
the FFT), we do not see any Zeeman splitting of the $\alpha$ nor
$\beta$ peaks. This is generally the case in metals, as the Zeeman
energy is small compared to the Fermi
energy\cite{CeIn3-Harrison,YRS-2008}. These results rule out a major
Fermi Surface reconstruction associated with the $\alpha$ and
$\beta$ Fermi surface sheets accompanying the metamagnetic
transition in
CeIrIn$_{5}$.\\

\begin{figure}
\resizebox{!}{0.8\textwidth}{\includegraphics{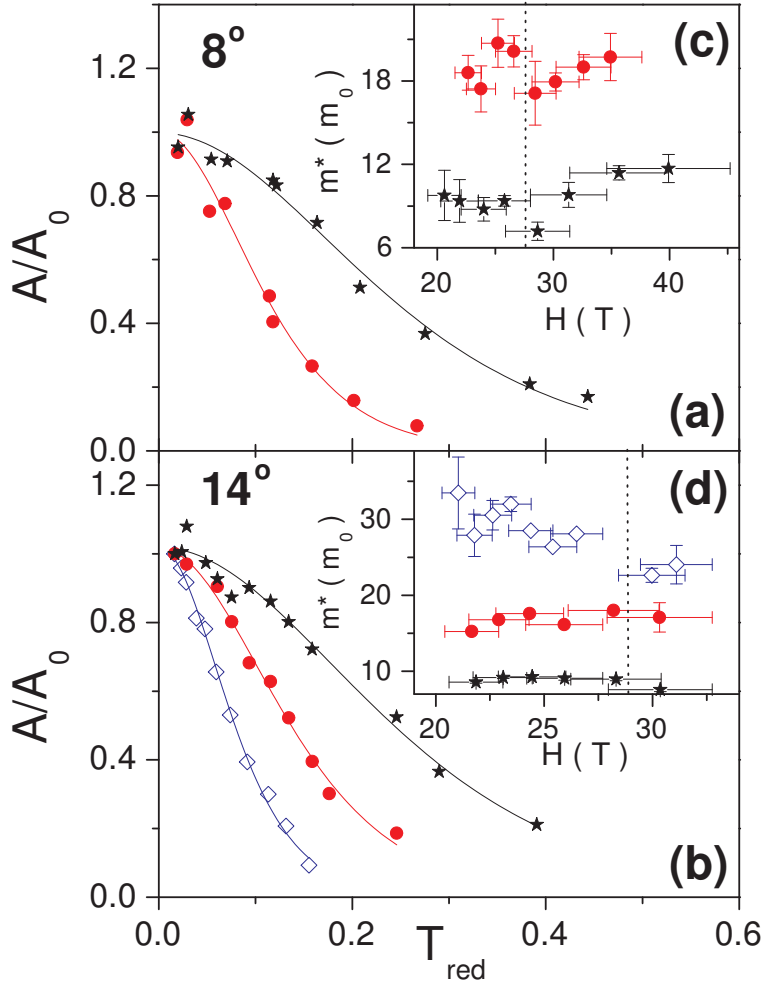}}
\caption{\label{Meff} Determination of effective masses from dHvA in
CeIrIn$_5$. (a) and (b) Normalized dHvA Amplitude, $A$, vs reduced
temperature, $T_{red}=\frac{\pi^2 k_B T}{\mu_B H}$, for orientations
of $\Theta=-8^{\circ}$ and $\Theta=14^{\circ}$ between $H$ and
[001], for the branches $\alpha_{3} (\star), \beta_{2} (\bullet),
\beta_{1} (\diamond)$. The amplitudes shown are from the FFT spectra
on a small field range below $H_m$. The solid lines are fits to data
with the Lifshitz-Kosevich (LK) formula (see text) with $A_{0}$ and
$m^{\ast}$ being adjustable parameters. The amplitudes shown are
normalized with respect to $A_{0}$, the $T=0$ value. (c) and (d)
Effective mass vs magnetic field for $\Theta=-8^{\circ}$ and
$\Theta=14^{\circ}$ determined from the LK fits. The dashed lines
correspond to the IEM transition field $H_m$. The vertical and
horizontal error bars represent the uncertainty in the fitting
parameter and the field range of the FFT respectively. The FFT is
performed on intervals of equal length in $H^{-1}$ centered around
the indicated fields, corresponding to 29 periods for each branch
for the 45 T data (a) and 23, 35 and 37 periods for $\alpha_{3},
\beta_{2}$ and $\beta_{1}$ for the 33T data (b).}
\end{figure}

\indent In the LK theory\cite{Shoenberg}, the effective cyclotron mass
is determined from the temperature dependence of the amplitude of the
dHvA oscillations. The dHvA amplitudes decrease with increasing
temperature due to the thermal broadening of the Landau levels
crossing the Fermi surface. We have determined the dHvA amplitudes by
performing a FFT on intervals of equal length in inverse magnetic
field $H^{-1}$ centered around a field value ranging between 20 and 33
T for the first data set ($\Theta=14^{o}$) and between 20 and 45 T for
the second data set ($\Theta= -8^{o}$).  Each interval contains a
fixed number of periods of a given dHvA frequency: 29 periods of
$\alpha_3$ and $\beta_2$ at $-8^{\circ}$ (45 T data set) and 23, 35,
and 37 periods of $\alpha_3$, $\beta_2$ and $\beta_1$ at $14^{\circ}$
(33 T data set). The integer number of periods in a given field
interval ensures no artificial broadening of the peaks with the
periodic boundary conditions used in the FFT.  The temperature
dependence of the dHvA amplitudes is then fit to the LK formula in
order to determine the effective masses for each branch at a given
field (see Ref.~\cite{Shoenberg}, p.60): $$A=A_0
\frac{xT}{sinh(xT)},$$ with $x=\frac{\alpha \frac{m^{\ast}}{m_e}}{H}$,
where $A$ is the dHvA amplitude, $A_0$ the $T=0$ amplitude, $m^{\ast}$
the effective mass, $m_e$ the bare electron mass, and $\alpha =
\frac{\pi^2 k_B}{\mu_B} = 14.69$ T/K. The data and fits shown in
Figs.~\ref{Meff}(a) and \ref{Meff}(b) are from FFT spectra on a field
range below $H_m(0)=28$ T. The field evolution of $m^{\ast}$ across
the metamagnetic transition is determined in this fashion for each
dHvA branch and is shown in Figs.~\ref{Meff}(c) and \ref{Meff}(d). The
cyclotron mass decreases with increasing field for the $\beta_1$
orbit, and is rather field independent for the other two orbits,
$\alpha_3$ and $\beta_2$.  In contrast to other metamagnetic
systems\cite{CeRu2Si2-dHvA1,CeRu2Si2-dHvA2,Sr327-dHvA}, no divergence
of $m^{\ast}$ is observed in any branch for fields near $H_{m}$,
within the finite field resolution of our procedure set by the field
interval of the FFTs. For the branches $\alpha_3$ and $\beta_2$ the
effective masses show no significant angular variation between
$\Theta=-8^{\circ}$ (Fig.~\ref{Meff}(c)) and $\Theta=14^{\circ}$
(Fig.~\ref{Meff}(d)). At $-8^{o}$, the $\beta_1$ orbit is not
resolved, possibly due to its large mass. The effective masses we
obtained are also in close agreement with the previously reported
masses for magnetic fields below $16$ T in the [001]
direction\cite{Ir115dHvA}.\\

\indent The evolution of the Fermi surface has been studied via dHvA
effect in a number of heavy fermion compounds across metamagnetic
transitions\cite{YRS-2008,
CeRu2Si2-dHvA1,CeRu2Si2-dHvA2,Julian-dHvA}. These studies have shown
that the IEM is in general accompanied by drastic changes both in
the effective mass and in the dHvA frequencies. This frequency
change has been ascribed to $f$-electron localization in
CeRu$_2$Si$_2$\cite{CeRu2Si2-dHvA1,CeRu2Si2-dHvA2} or to Zeeman
splitting (Fermi surface polarization) in UPt$_3$\cite{Julian-dHvA}.
In the light of these results, the absence of a substantial
frequency change in CeIrIn$_5$ is striking, despite the fact that
the field scale of $H_m$ is comparable to or larger than those of
CeRu$_2$Si$_2$ and UPt$_3$. This implies either a comparable or
larger Zeeman energy in CeIrIn$_5$. Recently, the IEM in CeIrIn$_5$
was attributed to a valence transition\cite{miyake} in which case we
would also expect a change in the Fermi Surface volume. Our data
appears to rule out this as a possibility. Thus, our data indicates
that neither the Zeeman splitting, and associated proximity to a van
Hove singularity, nor $f$-electron localization, or even valence
fluctuations are the driving mechanisms for IEM in CeIrIn$_5$.\\

\begin{figure}
\resizebox{!}{0.8\textwidth}{\includegraphics{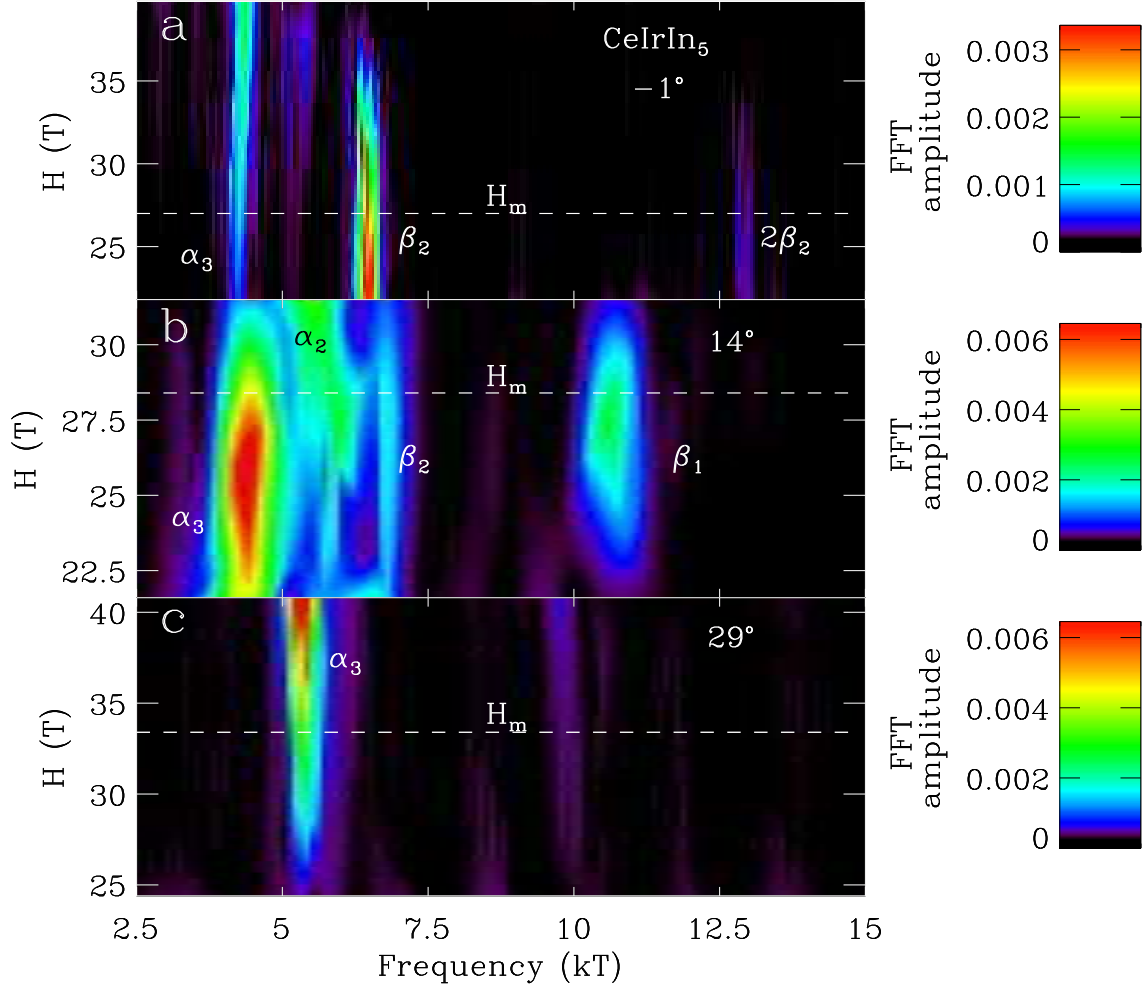}}
\caption{\label{FFTvsField} de Haas-van Alphen amplitude variation
near the metamagnetic critical field in CeIrIn$_5$. Contour plot of
the FFT Amplitude vs. frequency and magnetic field, $H$, at $T=40$
mK and (a) $\Theta=-1^{\circ}$, (b) 14$^{\circ}$, and (c)
29$^{\circ}$. The FFT spectra was obtained on a field interval
centered around the indicated fields, all intervals being of equal
length in $H^{-1}$, the length chosen to correspond to 29 periods
(panel a,c) and 23 periods (panel b) of $\alpha_3$.  The $\alpha_1$
and $\alpha_2$ orbits are not well resolved due to the smaller field
range of the FFT leading to broader peaks. The high frequency peak
in (a) is the second harmonic of the $\beta_2$ orbit.}
\end{figure}

\indent Despite the lack of drastic changes in neither the dHvA
frequencies nor effective masses, we observe distinct changes in the
dHvA signal near $H_m$. Namely, the metamagnetic transition is
accompanied by an overall damping of the dHvA oscillations. This can
be directly observed on the background subtracted torque signal of
Fig.~\ref{Mosc}. At $\Theta=-1^{\circ}$ (Fig.~\ref{Mosc}a), only the
lowest frequency $\alpha_3$ oscillations are observed above the
transition field $H_m=27.5T$, the $\beta_2$ oscillations being
completely suppressed in the field polarized state. At
$\Theta=29^{\circ}$ (Fig.~\ref{Mosc}b), where $\alpha_3$ is the only
frequency present in the entire field range, we see clearly that the
amplitude of the oscillations does not increase monotonically with
magnetic field but instead goes through a minimum around
$H_m=33.4T$. The field evolution of the FFT spectra for three
representative angles ($-1^{\circ}$, $14^{\circ}$ and $29^{\circ}$)
is depicted on the contour plots of the dHvA amplitude vs magnetic
field and frequency in Fig.~\ref{FFTvsField}. The frequencies of the
$\alpha_3$, $\beta_1$ and $\beta_2$ orbits being all field
independent, the corresponding contours stretch vertically. As
clearly seen in Fig.~\ref{FFTvsField}, the amplitude of the
$\beta_2$ orbit is completely suppressed at $-1^{\circ}$ in the
field polarized state above $H_m$. A similar trend is also seen for
the $\beta_1$ peaks at $14^{\circ}$ (the data is only available up
to 33 T at this angle). The $\beta_2$ peak at $14^{\circ}$ is not
easily distinguished from the neighboring $\alpha_1$, $\alpha_2$
peaks in the spectra due to the insufficient frequency resolution
for small field ranges used in the FFT. In contrast, the amplitude
of the $\alpha_3$ peak is largest in the field polarized state after
going through a
minimum at a field just above $H_m$.\\

\begin{figure}
\resizebox{!}{0.8\textwidth}{\includegraphics{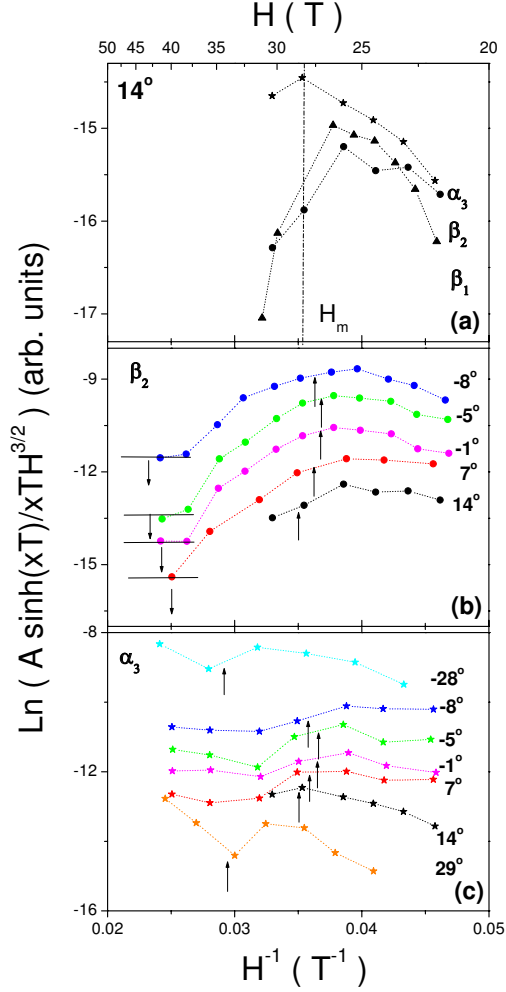}}
\caption{\label{Dingle} Dingle Temperature analysis in CeIrIn$_5$.
(a) Reduced amplitude $\ln(A\frac{sinh(xT)}{xT H^{\frac{3}{2}}})$
vs.\ inverse field $H^{-1}$ for $\Theta=14^o$ for the orbits
$\alpha_3$ ($\star$), $\beta_1$ ($\bigtriangleup$), $\beta_2$
($\bullet$). For convenience, the corresponding field range is also
indicated at the top of the figure. (b) Reduced amplitude vs inverse
field for $\beta_2$ orbit at all angles where this frequency is
resolved, angles ranging from $-8^{\circ}$ to $14^{\circ}$ as
indicated. The horizontal lines with arrow correspond to the noise
level of the FFT spectra giving an upper limit for the amplitudes at
the highest fields. (c) Reduced amplitude vs inverse field for
$\alpha_3$ orbit at various angles. The data for all angles is up to
45 T, except for $\Theta=14^{\circ}$, which is limited to 33 T. The
IEM transition field $H_m$ is indicated for each angle with a
vertical arrow.}
\end{figure}

\indent One can see this trend more clearly in the corresponding
Dingle plots, shown for the same angles ($-1^{\circ}$, $14^{\circ}$
and $29^{\circ}$) as well as for several others in
Fig.~\ref{Dingle}. Here the reduced dHvA amplitude as a function of
inverse magnetic field $H^{-1}$ is displayed. The reduced amplitude
is defined as $\ln(A\frac{sinh(xT)}{xT H^{\frac{3}{2}}})$, where $A$
is the FFT amplitude of each orbit.  According to the LK theory (see
Ref.~\cite{Shoenberg}, p.66), the reduced amplitude should be linear
in inverse field. As seen in Fig.~\ref{Dingle}a, this is not the
case in CeIrIn$_5$ and the amplitudes of the $\beta_1$ and $\beta_2$
oscillations are strongly suppressed in the field polarized state
for $H>H_m=29$ T at $14^{\circ}$. Moreover, the suppression of the
$\beta_2$ oscillations is not restricted to a particular angle but
is systematically observed for the entire range where this orbit is
resolved ($-8^{\circ} \leq \Theta \leq +14^{\circ}$), as shown in
Fig.~\ref{Dingle}b. In contrast, the amplitude of the $\alpha_3$
oscillations go through a minimum above the transition but recover
at higher fields, as shown in Fig.~\ref{Dingle}c for various angles
ranging from $-28^{\circ}$ to $+29^{\circ}$. The maximum in the
amplitudes of both $\alpha_3$ and $\beta_2$ moves to higher fields
(smaller $H^{-1}$) with increasing angle, following the same trend
as $H_m$ (see Figs.~\ref{Dingle}(b), \ref{Dingle}(c)). Thus, the
main result of our investigations is that there is a selective
damping of the dHvA oscillations in the field-polarized state, due
to the metamagnetic transition.

\indent In LK theory, the slope of the Dingle plot corresponds to
the Dingle temperature defined as $T_D=\frac{\hbar \tau^{-1}}{2\pi
k_{B}}$, with $\tau^{-1}$ the scattering rate on each orbit
describing the impurity broadening of the Landau levels at
low-$T$\cite{Shoenberg}. For $\alpha_3$ and $\beta_2$ orbits, Dingle
temperatures of 0.31 and 0.27 K were obtained from linear fits to
the reduced amplitudes in the low field state, for
$\Theta=-8^{\circ}$, using the effective masses determined
independently from the LK fits (see above). These correspond to a
mean free path of 112 nm for the $\alpha_3$ orbit and 172 nm for the
$\beta_2$ orbit which compares well to the value of 105 nm reported
previously\cite{Ir115dHvA}.  The scattering rate associated with
such values of $\ell$ is of the order of 3 THz, which is an order of
magnitude less than the scattering time estimated from $\rho_0$.
This is not surprising given the crude one band, one relaxation time
approximation of the Drude formula as well as the differences in the
effective cyclotron masses and that estimated from the Sommerfeld
model and $\gamma_0$ ($\sim 260 m_e$) used in Eq.~\ref{eq:drude}.
Nevertheless, the fact that the Dingle temperatures for both orbits
are $\sim 0.3$ K which is of the order of both T$_0$ and T$_{min}$
(see above) suggests that the transport is indeed in the diffusive
regime with predominant disorder scattering. For $\beta_1$, we could
not determine $T_D$ since the Dingle plot is not linear in the
inverse magnetic field, even in the low field state (see
Fig.~\ref{Dingle}a). Thus the scattering rate for this orbit, as
well as the effective mass (see above), appear to be somewhat field
dependent, leading to a breakdown of the Dingle analysis.\\

\indent The observed decrease in the amplitude of the dHvA
oscillations above $H_{m}$ for some of the observed branches is
unexpected for at least two reasons. First, the spin fluctuations
are suppressed in a field-polarized state which would indicate a
smaller scattering rate, and thus larger oscillation amplitudes.
Second, if one takes into account the magnetization in the dHvA
effect, in fact the oscillations are in $B^{-1}$ rather than in
$H^{-1}$, one would also expect a positive feedback with the larger
internal field giving rise to a larger amplitude of the
oscillations. A reduction of the amplitude for all of the orbits
near $H_m$ may indicate an increased scattering rate due to electron
scattering from magnetic fluctuations
which are expected to be maximized at the phase transition.\\

\indent There are several reasons for a decreased dHvA amplitude at
high magnetic fields that are understood from extensive dHvA
experiments in simple metals. One such possibility for a decreased
dHvA amplitude is the so-called  magnetic interaction leading to
phase smearing\cite{Shoenberg}, corresponding to the suppression of
the overall signal due to electrons out-of-phase with one another in
different parts of the sample. The underlying assumption is that the
sample is either inhomogeneous in real space or that there is an
inhomogeneous internal field distribution. In fact, if the
magnetization is large then there will be an internal field
inhomogeneity in the sample due to the shape dependent demagnetizing
factor.  Such an inhomogeneity would cause the dHvA amplitude to be
smaller than when the field is homogeneous throughout the sample.
This internal field correction is relevant when the magnetization of
the sample is large. From the magnetization measured with a
commercial SQUID at $1.8$ K at an applied field of $7$ T, we have
extrapolated the volume magnetization in CeIrIn$_5$ to be about
$326$ G at $30$ T (without taking into account the increase in M due
to the metamagnetic transition), which is 3 orders of magnitude
smaller than the external field. Unless the sample itself is in an
intrinsically inhomogeneous state, we are safe in assuming that $B
\approx H$ is nearly homogeneous. To the best of our knowledge, no
evidence for sample inhomogeneity is reported in the undoped, highly
stoichiometric, high quality single crystals of CeIrIn$_5$. Also it
is not clear how this effect would lead to preferential damping of
the $\beta$ orbits, nor to the damping only in the field polarized
state. Thus phase smearing is unlikely to be the cause of the
observed damping.\\

\indent A second possibility for a non-linear Dingle plot is the
magnetic breakdown for the $\beta$ orbits. Magnetic breakdown
happens when the applied field is strong enough that the electrons
tunnel from one piece of the Fermi Surface to another, leading to
the observation of large dHvA frequencies\cite{Shoenberg}. Since we
do not see any new frequencies appearing in the field polarized
state, this is unlikely in our case. Finally, the dHvA amplitude can
be suppressed due to the spin reduction factor, namely the
destructive interference between the Zeeman split up- and down-spin
electrons. Although we cannot resolve this splitting in the dHvA
frequencies, such a polarization would lead to a decrease of the
dHvA amplitudes. The spin reduction factor\cite{Shoenberg}, $R_s =
cos(\frac{\pi g m*}{2m})$, can be estimated for various orbits using
the cyclotron effective masses determined above and with a Lande
factor $g=2$, yielding the same reduction factor for $\alpha$ and
$\beta$ orbits. Thus we can rule out Zeeman splitting as the cause
of selective damping of dHvA oscillations above $H_m$. \\

\indent Thus, one is lead to conclude that there has to be an
additional damping mechanism beyond those typically considered
(thermal and  disorder broadening or Zeeman splitting) in the LK
theory. Perhaps the most noteworthy aspect of our data is that the
damping is selective: the heavier $\beta-$orbits are more strongly
damped than the lighter $\alpha-$orbits, being totally suppressed in
the field-polarized state. Although the cyclotron mass of all
observed orbits does not change appreciably across the transition, a
lighter heavy fermion ground state in the high field state would
result due to this selective suppression of heavy quasiparticles
from the heavier $\beta$-sheet. It is also interesting to note that
the $\beta$ orbits are associated with the same hole band for which
the smallest of the four hybridization gaps, namely $\Delta_1$, has
been assigned in optical conductivity measurements\cite{burch}.
Although the interpretation of optical conductivity data may not be
unique, it points to the same trend in the evolution of this gap
across the CeMIn$_5$ family as the systematic changes found in other
characteristic parameters such as $\gamma$ and $T_c$\cite{burch}.\\

\indent The decreasing dHvA amplitudes for the $\beta-$orbits above
$H_m$ are also difficult to reconcile with the simultaneous drop in
$\rho$. Assuming that $\rho$ is determined by parallel conduction
channels associated with each Landau orbit, and  since the heavier
$\beta-$orbits (with smaller conductivity) are preferentially damped
above $H_m$, this would lead to an increased $\rho$, making
necessary a simultaneous enhancement of the mean free path $\ell$ of
the light carriers in order to have a decreased $\rho$. Thus, the
transport is most likely dominated by the lighter electrons on the
$\alpha$ sheet that are strongly scattered near the transition but
are otherwise intact in the field-polarized state.  This is a
possible explanation of the apparent paradox of a simultaneous
decrease in both $\rho$ and the dHvA amplitudes across the IEM
transition.

\section{Conclusion}

 \indent In conclusion, our comparative study of
$\rho$ and dHvA oscillations is suggestive of an unusual
metamagnetic transition in CeIrIn$_5$. The observed deviations from
quadratic behavior in $\rho$ vs $T$ above as well as below $H_m$,
may well correspond to a non-Fermi Liquid regime extending over a
wide range of magnetic field in the phase diagram. In the
field-polarized state, above 36 T, a Fermi liquid regime is
recovered. Moreover, an upturn is observed in the $T-$dependence of
$\rho$ that is somewhat different from what is reported in
Sr$_3$Ru$_2$O$_7$\cite{Sr327-upturn,Sr327-science07}. The $\rho(T)$
that we measure may correspond to quantum corrections that occur in
a diffusively conducting strongly correlated metal. A detailed
analysis of the dHvA oscillations shows that while the frequencies
and effective masses are not significantly changed in the vicinity
of the transition, the amplitude of some of the orbits are
unexpectedly suppressed in the high field state. One possible way to
reconcile the anomalous damping of the dHvA oscillations with the
pronounced drop in the magnetoresistance is that the heavy holes are
selectively damped across the IEM transition and that the transport
is dominated by the lighter electrons. These results represent a
challenge to our current understanding of IEM and call for further
investigation as to what can cause such a change in the
magnetization, the metamagnetic transition, beyond the mechanisms
thus far explored, none of which can adequately account for the
observed response in CeIrIn$_5$.

\begin{acknowledgments}
We acknowledge fruitful discussions with B.Binz, I. Vekhter, F.
Ronning, P. Adams, A. Chubukov, P. Schlottmann, C. Varma and
D.Pines. Work at Los Alamos was performed under the auspices of the
U.S. Department of Energy. A portion of this work was performed at
the National High Magnetic Field Laboratory, with support from the
NSF Cooperative Agreement No. DMR-0084173, by the State of Florida,
and by the DOE. Work at LSU was supported by NSF under grant no.
DMR-0804376.
\end{acknowledgments}

%\bibliography{CeIrIn5_2ndedition}

\end{document}